\documentclass[preprint2]{aastex}
\begin{document}

\title{\large{\rm{ON THE FORM OF THE SPITZER LEAVITT LAW AND ITS DEPENDENCE ON METALLICITY}}}

\author{\small D. Majaess$^1$, D. Turner$^2$, W. Gieren$^3$}
\affil{$^1${\footnotesize Halifax, Nova Scotia, Canada.}}
\affil{$^2${\footnotesize Department of Astronomy and Physics, Saint Mary's University, Halifax, NS B3H 3C3, Canada.}}
\affil{$^3${\footnotesize Departamento de Astronom\'ia, Universidad de Concepci\'on, Casilla 160-C, Concepci\'on, Chile.}}
\email{dmajaess@cygnus.smu.ca}

\begin{abstract}
The form and metallicity-dependence of Spitzer mid-infrared Cepheid relations are a source of debate.  Consequently, Spitzer 3.6 and 4.5 $\mu$m period-magnitude and period-color diagrams were re-examined via robust routines, thus providing the reader an alternative interpretation to consider.  The relations (nearly mean-magnitude) appear non-linear over an extensive baseline ($0.45<\log{P_0}<2.0$), particularly the period-color trend, which to first-order follows constant (3.6-4.5) color for shorter-period Cepheids and may transition into a bluer convex trough at longer-periods.   The period-magnitude functions can be described by polynomials (e.g., $\left[ 3.6 \mu m \right] =K_0 -(3.071\pm0.059) \log{P_0}-(0.120\pm0.032)\log{P_0}^2$), and Cepheid distances computed using 3.6 and 4.5 $\mu$m relations agree and the latter provides a first-order consistency check (CO sampled at 4.5 $\mu m$ does not seriously compromise those distances).  The period-magnitude relations appear relatively insensitive to metallicity variations ($\rm{[Fe/H]}\sim0$ to -0.75), a conclusion inferred partly from comparing galaxy distances established from those relations and NED-D ($n>700$), yet a solid conclusion awaits comprehensive mid-infrared observations for metal-poor Cepheids in IC 1613 ($\rm{[Fe/H]}\sim -1$).   The Cepheid-based distances were corrected for dust obscuration using a new ratio (i.e., $A_{3.6}/E_{B-V}=0.18\pm0.06$) deduced from GLIMPSE (Spitzer) data.  
\end{abstract}

\section{{\rm \footnotesize INTRODUCTION}}
Spitzer mid-infrared observations of Cepheid variables have been employed to constrain stellar mass-loss and anchor the cosmic distance scale \citep[e.g.,][]{ne09,mo12}.  Regarding the latter, the principal aim is to establish firmer constraints on the Hubble constant and cosmological models \citep{fr11}.  However, ambiguities linger concerning the form and metallicity-dependence of mid-infrared Cepheid relations.

The impetus for obtaining Spitzer observations stems partly\footnote{See \citet[][their \S 2]{fr11}.} from the diminished impact of dust extinction on infrared-based distances ($A_{3.6}/A_V\sim0.06$, \S \ref{s-ext}).  Crucially, uncertainties tied to neglecting variations in the extinction law are reduced. The extinction law varies throughout the Galaxy, and a $20$\% uncertainty ($\sigma_{R_V}/R_V\sim0.2$) merely contributes $\sigma_{\mu_0}\sim0^{m}.01$ to the mid-infrared distance modulus (assuming $E_{B-V}=0.10\pm0.03$).  Conversely, variations in the extinction law can significantly impact Cepheid and star cluster distances that rely solely on shorter-wavelength data \citep[e.g.,][and discussion therein]{tu12,ca12}.

The aforementioned points reiterate the importance of Spitzer observations.  However, a partial account provided below highlights the diverse opinions expressed concerning the form and metallicity-dependence of mid-infrared Cepheid relations.  \citet{nk08}, \citet{ma09}, and \citet{mar10} cited different slopes for linear Spitzer period-magnitude relations, while \citet{ne09} suggested that function is non-linear \citep[see also][]{nk10}.   For example, \citet{nk08} obtained $\alpha=-3.26$ for the slope of the 3.6 $\mu$m relation using the OGLE\footnote{Optical Gravitational Lensing Experiment \citep[e.g.,][]{so08,so10}.} and SAGE\footnote{Surveying the Agents of a Galaxy's Evolution \citep{me06,go11}.} datasets ($0.5< \log{P} <1.7$), whereas \citet{ma09} favored a steeper slope inferred by correlating the \citet{pe04} and SAGE catalogs  ($\log{P}>0.8$).  A shallower slope may arise partly from the inclusion of shorter-period Cepheids that could suffer increased photometric contamination, particularly for distant targets \citep[e.g.,][]{mac06,ma09,maj13}.  However, \citet{nk10} deduced analogous period-magnitude relations for relatively nearby Cepheids in the SMC that occupy low and high density regions.

\begin{figure*}[!t]
\begin{center}
\includegraphics[width=17cm]{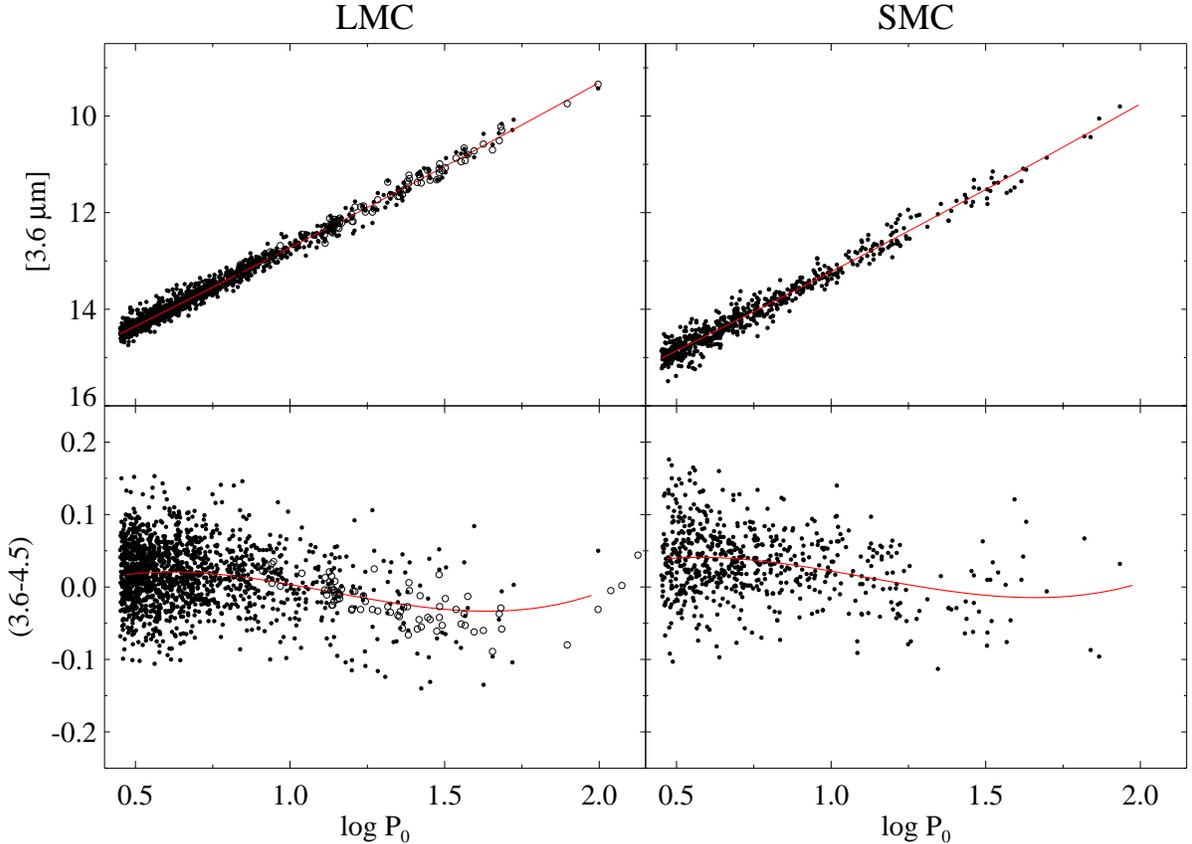} 
\caption{\small{Spitzer period-magnitude and period-color diagrams for Cepheids in the Magellanic Clouds.  The (nearly mean-magnitude) data are described by non-linear trends over an extended baseline ($0.45<\log{P_0}<2.0$), particularly the period-color relation.  Solid lines convey mean fits, and the data shown were cleaned of outliers via an iterative $\sigma$-clip.  Filled circles represent SAGE photometry, while open circles are observations from \citet{sc11}.}}
\label{fig-lmcpm}
\label{fig-lmcpc}
\end{center}
\end{figure*}

\citet{mar10} and \citet{ng12a} found certain models produced Spitzer period-magnitude and period-color relations that did not match observations spanning an extended abundance baseline \citep[see also][and discussion therein]{ng12b}. Observations imply that the same slope characterizes 3.6 and 4.5 $\mu$m period-magnitude relations for SMC and LMC Cepheids \citep{nk10}, which exhibit mean abundances of $\rm{[Fe/H]}\sim-0.75,-0.33$ accordingly \citep{lu98,mo06}.  \citet{mar10} attempted to extend the baseline to include Galactic Cepheids ($\rm{[Fe/H]}\sim0$), but the effort was marred by inconsistencies associated with the available calibrating data \citep[see also][and discussion therein]{nk10}.  \citet{hg74} and \citet{mar10} noted that the 4.5 $\mu$m passband samples CO absorption features, which presumably alter the Spitzer Cepheid relations and introduce a metallicity effect.  \citet{sc11} confirmed the former finding by demonstrating that cooler long-period Cepheids exhibit colors that are modulated by the absorption and disassociation of CO, whereas their hotter short-period counterparts are less affected since the molecule is gradually disassociated. \citet{fm11} estimated that 3.6 and 4.5 $\mu$m Cepheid data exhibit a metallicity dependence of $\gamma=-0.39\pm0.16$ and $-0.25\pm0.18$ mag dex$^{-1}$, respectively.  The former estimate was revised by \citet{fr11} to $\gamma=-0.09\pm0.29$ mag dex$^{-1}$.  Those results are tied partly to published abundance estimates for individual LMC Cepheids, yet it is unclear whether systematic effects pervade such individual determinations,\footnote{The \citet{fm11} analysis, which relies on published abundance estimates found in the literature for individual Cepheids, implies that the $BV$ Wesenheit function displays a nearly negligible dependence on metallicity, whereas certain researchers suggest otherwise \citep{ta03,maj09,bo10}.} although a mean metallicity derived from an entire sample appears reliable \citep{lu98,ro08}.  

In this study the form and metallicity dependence of Spitzer period-magnitude and period-color relations are re-investigated. The relation linking the total extinction at 3.6 $\mu$m ($A_{3.6}$) to the optical color-excess ($E_{B-V}$) is likewise derived from Spitzer data, as the correlation is necessary for computing Cepheid distances using the period-magnitude relations derived.

\section{{\rm \footnotesize ANALYSIS}}
\subsection{{\rm \footnotesize 3.6 AND 4.5 $\mu$m PERIOD-MAGNITUDE RELATIONS}}
3.6 and 4.5 $\mu$m period-magnitude relations are now inferred from LMC and SMC Cepheids.  The LMC sample was assembled by correlating HV\footnote{Harvard Variable.} Cepheids tabulated by \citet{ma85} and \citet{pe04} with SAGE (Spitzer) photometry (a mean of random-phase data).  Fundamental mode Cepheids cataloged by OGLE were likewise added to the sample.  For the SMC analysis, the \citet{ma85} and OGLE catalogs were paired with SAGE photometry.  A period cutoff of $\log{P_0}>0.45$ was imposed to: mitigate biases stemming from a reputed change in slope near $\log{P}\sim0.4$ \citep{nk10}, to ensure magnitude uncertainties were satisfactory, to reduce the effects of photometric contamination, and to ensure that brighter Cepheids on the blue-edge of the instability strip were not preferentially sampled \citep[\citealt{sa88}, see also Appendix A in][]{fr01}.  The $\log{P}<2.0$ limit was adopted to avoid the inclusion of Leavitt variables \citep{gr85}, which may adhere to a separate slope \citep[][their Fig.~3]{pe04}. 

The nearly mean-magnitude period-color (3.6-4.5) relation examined in \S \ref{s-pcr} is characterized by a higher-order polynomial over the baseline examined, thus hinting that the period-magnitude relations are non-linear.   A single linear period-magnitude function does not adequately describe both the short and long-period domains \citep[see also][]{ne09}. Longer-period Cepheids ($\log{P}>1.45$) displayed a $\Delta[3.6 \mu m] \gtrsim 0^{m}.06$ offset relative to a linear fit, whereby the latter fit yields values systematically too faint.  Longer-period Cepheids are  important granted the stars are preferentially sampled within remote galaxies in the Hubble flow.  Efforts toward achieving precision cosmology ($\sigma_{H_0}/H_0<3$ \%) hinge on reducing sources of uncertainty for such stars, especially since certain Cepheid-based determinations of $H_0$ exhibit tension with the new Planck results \citep[][and discussion therein]{fr12,tr12}.    

Robust fitting routines were employed since the method of least-squares is acutely sensitive to outliers,\footnote{Certain outliers may be Cepheids exibiting high mass-loss \citep{ne09}.} whereas the former techniques are less affected by spurious data and more apt to identify the underlying trend.  The robust routine applied seeks to minimize the sum of the absolute deviations, rather than the sum of the squared deviations.  The routine was used in concert with a $4\sigma$ clip, twice iterated.  That conservative procedure was followed because \citet{ma09} suggested that differences between the various determinations for the period-magnitude relation may arise partly from aggressive clipping/iterations.  

The following functional forms were deduced and characterize LMC Cepheids across an extensive baseline ($0.45<\log{P_0}<2.0$):
\begin{eqnarray}
\label{eq-lmcpm}
\nonumber
\left[ 3.6 \mu m \right] =(15.917 \pm 0.026) -(3.077\pm0.064) \log{P_0} 
\\ \nonumber -(0.114 \pm0.036)\log{P_0}^2
\nonumber \\
\left[ 4.5 \mu m \right] =(15.890 \pm 0.026) -(3.084\pm0.065) \log{P_0} 
\\ \nonumber  -(0.082\pm0.036)\log{P_0}^2
\end{eqnarray}
For SMC Cepheids the corresponding relations are: 
\begin{eqnarray}
\label{eq-smcpm}
\nonumber
\left[ 3.6 \mu m \right] =(16.411 \pm 0.063) -(3.041\pm0.143) \log{P_0} 
\\ \nonumber -(0.145\pm0.074)\log{P_0}^2
\nonumber \\
\left[ 4.5 \mu m \right] =(16.382 \pm 0.061) -(3.076\pm0.140) \log{P_0} 
\\ \nonumber -(0.104\pm0.072)\log{P_0}^2
\end{eqnarray}
The magnitude of the squared term increases slightly if longer-period Cepheids are given more weight, and the uncertainties subsequently diminish. Yet a single quadratic expression is preferred over two linear relations since the statistical weight of the entire Cepheid demographic can be exploited to reduce uncertainties. Relatively nearby short-period Cepheids detected in the Magellanic Clouds outnumber their long-period counterparts, as low-mass metal-poor stars cross the instability strip with an extended blue-loop \citep{be85}.  The initial mass function implies that low-mass objects are more numerous, which in concert with the aforementioned trend, explains the shift in the distribution of Magellanic Cloud Cepheids toward shorter-periods.\footnote{There exists a period-mass-luminosity relation \citep[e.g.,][]{tu12b}.}    

\subsection{{\rm \footnotesize METALLICITY EFFECTS AT 3.6 AND 4.5 $\mu$m}}
The coefficients tied to the pulsation period in Eqns.~\ref{eq-lmcpm} and \ref{eq-smcpm} agree to within the uncertainties.  Thus the slopes for the 3.6 and 4.5 $\mu$m period-magnitude relations are comparatively insensitive to metallicity variations from $\rm{[Fe/H]}\sim-0.33$ to -0.75 \citep[see also][]{nk10}.  The 3.6 and 4.5 $\mu$m zero-points and coefficients for a given Magellanic Cloud agree to within the uncertainties, thus indicating that distances established from the latter passband can serve as a first-order consistency check \citep[CO is sampled in the 4.5 $\mu$m passband, e.g.,][]{hg74}.  

The impact of metallicity can likewise be assessed by comparing NED-D\footnote{NASA/IPAC Extragalactic Database Master List of Galaxy Distances \citep{sm11}.} and Spitzer-based Cepheid distances.  Relative offsets between the SMC/LMC, and the SMC/Milky Way, are now evaluated.  The latter comparison provides a desirably large abundance baseline ($\Delta \rm{[Fe/H]}\sim-0.75$).  Cepheid distances were calculated via the following relations derived from a weighted-mean of Eqns.~\ref{eq-lmcpm} and \ref{eq-smcpm}:
\begin{eqnarray}
\label{eq-smclmc}
\nonumber
\left[ 3.6 \mu m \right] =K_0 -(3.071\pm0.059) \log{P_0}
\\ \nonumber -(0.120\pm0.032)\log{P_0}^2
\nonumber \\
\left[ 4.5 \mu m \right] =K_0 -(3.083\pm0.059) \log{P_0} 
\\ \nonumber -(0.086\pm0.032)\log{P_0}^2
\end{eqnarray}
The offset between Cepheids in the SMC/LMC is $\Delta {K_{0,3.6}}=0.501\pm0.068$, while the differential for the SMC/Galaxy is $\Delta {K_{0,3.6}}=18.938\pm0.077$.  Marginal extinction corrections were applied to the Cepheid distances (\S \ref{s-ext}), and the Galactic calibration used is discussed below.  The corresponding 4.5 $\mu$m results are $\Delta {K_{0,4.5}}=0.480\pm0.066$ and $18.921\pm0.075$, respectively. The values compare favorably to means tabulated from NED-D data ($n>700$), namely $\Delta {K_0}=0.450\pm0.012$ and $18.930\pm0.011$.   The results imply values of $\gamma_{3.6}\sim-0.10\pm0.10,-0.01\pm0.06$ and $\gamma_{4.5}\sim-0.06\pm0.10,0.01\pm0.06$ mag dex$^{-1}$.  The associated uncertainties underscore the need for continued research, and formal uncertainties cited throughout the analysis may underestimate the true uncertainties. 

\begin{figure*}[!t]
\begin{center}
\includegraphics[width=17.2cm]{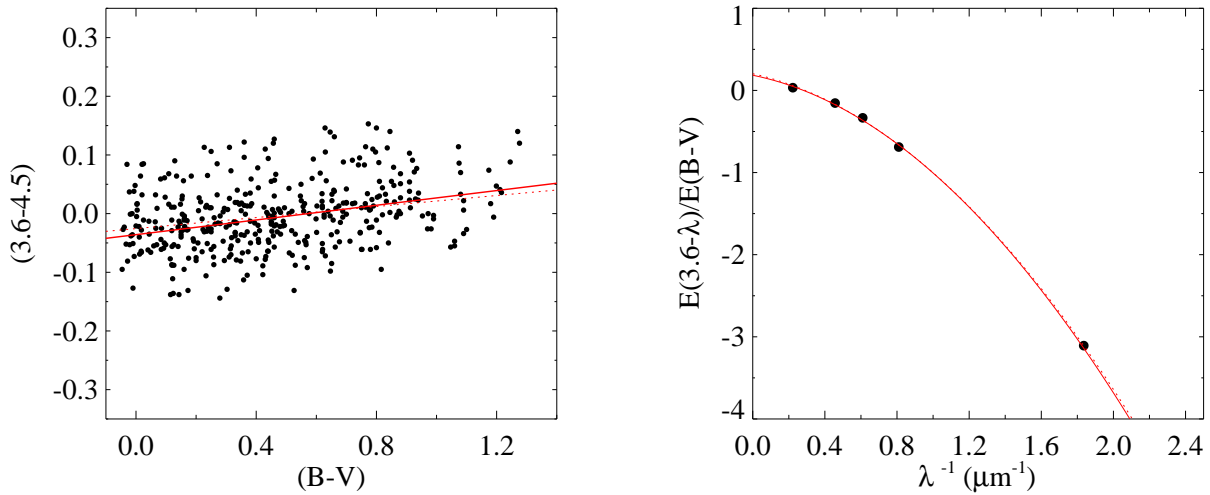} 
\caption{\small{Left, a linear fit yields the ratio $E_{3.6-4.5}/E_{B-V}=0.06\pm0.01$ (solid red line).  Right, a polynomial fit to data including Spitzer (GLIMPSE) observations yields $A_{3.6}/E_{B-V}=0.18\pm0.06$ (solid red line).  The ratios are used to correct Spitzer-based Cepheid distances for dust extinction, thus stymieing the propagation of a systematic uncertainty into the distance scale.  The dashed lines were inferred from parameters adopted by \citet[][see references therein]{mo12}.}}
\label{fig-ext}
\end{center}
\end{figure*}

\textit{In sum}, the analyses imply that distances inferred from the 3.6 and 4.5 $\mu$m period-magnitude relations are relatively insensitive to variations in metallicity ($\rm{[Fe/H]} \sim 0$ to -0.75).  However, a comparison including IC 1613 is required since Cepheids in that galaxy ($\rm{[Fe/H]}\sim-1$) are more metal-poor than their SMC counterparts.  Reliable Spitzer data are merely readily available for five $\log{P}<2$ Cepheids in IC 1613 \citep{fr09}, and a comparison at present may be premature given: those small statistics, the potential for inadequate instability strip filling, and concerns regarding photometric contamination expressed in the literature \citep{fr09}.  \citet[][their Table 1]{fr11} plan to obtain comprehensive mid-infrared observations for additional Cepheids in IC 1613.
 
The Galactic mid-infrared data were adopted verbatim from \citet{mo12}, who presented complete lightcurves for numerous Galactic Cepheids. \citet{mo12} paired their new photometry with IRSB,\footnote{Infrared Surface Brightness technique \citep{fg97}.} parallax, and cluster Cepheid data to establish a Galactic calibration \citep[e.g.,][]{be07,fo07,tu10}. That calibration consists of Cepheids with near solar abundances ($\rm{[Fe/H]}\sim0$), and is hence important for assessing the impact of metallicity.  However, the Galactic calibration may presently be the most difficult to construct, since the constituent Cepheids exhibit sizable differential reddening and do not lie at a common distance.  The Galactic calibration currently suffers from a lack of solid long-period calibrators, and parameters for cluster Cepheids are being contested and revised \citep[e.g.,][]{an12,maj12}.  The IRSB parameters for long-period Cepheids rely directly on the $p$-factor, the form of which is actively debated given its importance \citep{gi05,st11,ng12c,ne12}.\footnote{Gieren et al.~(2013, in preparation) aim to constrain the $p$-factor using LMC Cepheids in double-lined eclipsing binary systems \citep{pi10,pi11}.}   Nevertheless, the resulting non-linear fits  (Eqn.~\ref{eq-smclmc}) to the \citet{mo12} data\footnote{The absolute magnitudes used were an averaged subset from IRSB, cluster Cepheid, and the \citet{be07} parallax results \citep[see][]{mo12}.} yield: $K_{0,3.6}=-2.545\pm0.045$ and $K_{0,4.5}=-2.551\pm0.043$.   The results were derived by adopting the $\log{P_0}$ coefficient established from Magellanic Cloud Cepheids (Eqn.~\ref{eq-smclmc}), while permitting $\log{P_0}^2$ and $K_{0}$ to vary.  The aim was to examine whether the resulting coefficients for $\log{P_0}^2$ match those determined for Magellanic Cloud Cepheids (Eqn.~\ref{eq-smclmc}).  The results agree to first-order, namely $-0.10$ (3.6 $\mu$m) and $-0.05$ (4.5 $\mu$m).  However, those determinations are merely suggestive owing to the lack of numerous solid long-period calibrators.

\subsubsection{{\rm \footnotesize COMPARISON OF THE LMC DATA}}
The \citet{ma09} LMC sample results from a correlation of the \citet{pe04} and SAGE catalogs (a mean of random-phase data), whereas \citet{sc11} subsequently obtained complete mid-infrared lightcurve coverage for the Cepheids as part of the Carnegie Hubble Project.  Eqn.~\ref{eq-smclmc} is now applied to the LMC data of \citet{ma09} and \citet{sc11}.    The period coefficients remained fixed, and the zero-point was allowed to vary.  Zero-points of $K_{0,3.6}=15.939\pm0.030$ and $15.911\pm0.012$ were derived accordingly.  The results agree to within the formal uncertainties, and a visual inspection reveals a reliable fit (Fig.~\ref{fig-lmcpm}).  A least-squares solution was obtained since extreme outliers are absent.  

The \citet{sc11} data are tied to an independent photometric standardization, whereas the \citet{ma09}, \citet{nk08}, and Fig.~\ref{fig-lmcpm} samples rely on SAGE photometry. A comparison between the SAGE and \citet[][S11]{sc11} 3.6 $\mu$m photometry reveals a color offset ($\Delta [3.6]_{\rm S11-SAGE}\sim-0.9\times (3.6-4.5)_{\rm SAGE}$).  The offset may stem from uncertainties in the photometric standardization, but additional research is ultimately needed to identify the source.  
  
\subsection{{\rm \footnotesize PERIOD-COLOR RELATION}}
\label{s-pcr}
Period-color data for LMC and SMC Cepheids were examined (Fig.~\ref{fig-lmcpc}). The observations are characterized by a non-linear fit over the period baseline, whereby shorter-period Cepheids adhere to a near constant color and a convex trough may describe the longer-period domain.\footnote{The period-color relation is nearly constant across the entire period baseline when sampled at the hottest pulsation phase \citep[][their Fig.~9]{mo12}.} Those trends can be explained in part by CO \citep[see also][]{hg74,mar10,sc11,mo12}.  Fig.~\ref{fig-lmcpc} indicates that for shorter-period Cepheids the color is nearly constant as the temperature is sufficiently high to disassociate most CO.  For longer-period (cooler) Cepheids CO absorption occurs, and the molecule is gradually disassociated in increasing amounts near the pulsation phase corresponding to temperature maximum.  The period-color trend exhibits inflection points near $\log{P}\sim0.75$, $\log{P}\sim1.75$, and potentially $\log{P}>2.2$.  Additional analyses are warranted.   

A mean color inferred from shorter-period Cepheids near the constant part of the trend yields $(3.6-4.5)=0.023,0.039$ for the LMC and SMC, respectively. The optical to mid-infrared color-excess ratio derived in \S \ref{s-ext} implies intrinsic colors of $(3.6-4.5)_0=0.015,0.033$ (LMC/SMC).  Galactic Cepheids \citep{mo12} exhibit $(3.6-4.5)_0=-0.022$, as deduced from a visual match to intermediate-period Cepheids in the LMC.  That procedure was followed as the Spitzer Galactic sample does not presently contain numerous shorter-period Cepheids, and their longer-period counterparts can exhibit rather uncertain reddening estimates.\footnote{Turner (2013-14, in preparation) is reworking the Galactic cluster Cepheid calibration with the aim of improving intrinsic color and distance estimates.}

Offsets between the intrinsic colors of the three galaxies are likely insignificant owing to uncertainties arising from an inhomogeneous photometric standardization, and extinction corrections.  \citet{mo12} note that the systematic error associated with Spitzer photometry for Galactic Cepheids may be $\sim0^{m}.016$ in each passband (3.6 and 4.5 $\mu$m), and a similar uncertainty exists for the Magellanic Cloud photometry.  More sizable photometric offsets can exist in optical surveys of star clusters \citep[][their Table 3 for NGC 188]{st04} and the Magellanic Clouds (Majaess et al.~2013-14, in preparation).  Alternatively, comprehensive mid-infrared observations of Cepheids in IC 1613 will dictate whether metallicity offsets the period-color relation ($\gamma_{3.6-4.5}\sim-0.07$ mag dex$^{-1}+ ...$), as metal-poor Cepheids in IC 1613 are predicted to be intrinsically redder than their SMC, LMC, and Galactic counterparts \citep[][their Fig.~10]{mo12}.  

\subsection{{\rm \footnotesize EXTINCTION LAW}}
\label{s-ext}
Spitzer period-magnitude relations provide Cepheid distances once corrected for dust extinction.  Most extragalactic Cepheids observed for the HST key project to measure $H_0$ are reddened by $E_{B-V}\sim0.1$ \citep[e.g., Fig.~6 in][]{maj10}, and thus the resulting mid-infrared extinction is marginal.  There are exceptions such as the Cepheids in NGC 6822 \citep{gi06} and Centaurus A \citep{fe07}, which exhibit comparatively larger reddenings. 

The ratios $A_{3.6}/E_{B-V}$ and $E_{3.6-4.5}/E_{B-V}$ are often used since $E_{B-V}$ may be known \textit{a priori} from other observations.  Those ratios are now derived from Spitzer (GLIMPSE\footnote{The Galactic Legacy Infrared Mid-Plane Survey Extraordinaire. GLIMPSE surveyed a portion of the 4$^{\rm th}$ Galactic quadrant \citep{be03}.}) data for O-type stars, which exhibit comparable intrinsic $(B-V)_0$ colors and sizable reddenings.  O-type stars in the \citet{sk13} catalog were correlated with 2MASS, Spitzer, and optical photometry.  A linear relation between $(B-V)$ and $(3.6-4.5)$ is displayed in Fig.~\ref{fig-ext}, and defines $E_{3.6-4.5}/E_{B-V}$.  A robust fit applied using an iterative $\sigma$-clip yielded $E_{3.6-4.5}/E_{B-V}=0.06\pm0.01$.  

To determine $A_{3.6}/E_{B-V}$, color ratios ($E_{3.6-\lambda}/E_{B-V}$) were plotted as a function of $\lambda ^{-1}$ and extrapolated to $\lambda\to\infty$.  Intrinsic optical and near-infrared colors were adopted from \citet[][and references therein]{tu89} and \citet{sl09}.  Intrinsic mid-infrared colors were derived from the relevant color ratios (e.g., $(3.6-J)$ versus $(B-V)$ was extrapolated to the mean $(B-V)_0$ adopted for an O-type star).  Although O-type stars are advantageous given the reasons cited above, such young stars often occupy regions displaying PAH emission, and thus the 8.0 and 5.8 $\mu$m data were bypassed to mitigate concerns regarding contamination.  Fig.~\ref{fig-ext} displays the extinction law diagram, whereby a polynomial fit to the data yields $A_{3.6}/E_{B-V}=0.18\pm0.06$.  The ratios derived are consistent with values adopted by \citet[][see references therein]{mo12}.  

Lastly, extinction laws vary throughout the Galaxy \citep[e.g.,][and references therein]{ca12,na12}, and appear anomalous for certain regions in the 4$^{\rm th}$ Galactic quadrant sampled by GLIMPSE.  However, altering the ratio to account for variations in the extinction law does not significantly impact estimated mid-infrared distances, hence a motivation for monitoring Cepheids with Spitzer.   Applying the ratio derived above to the mean LMC reddening ($E_{B-V}\sim0.14$) yields $A_{3.6}\sim0^{m}.025$.  

\section{{\rm \footnotesize CONCLUSION}}
Spitzer 3.6 and 4.5 $\mu$m period-magnitude and period-color relations (nearly mean-magnitude) were re-investigated (Fig.~\ref{fig-lmcpm}).  The LMC and SMC period-magnitude functions, and particularly the period-color trend, appear non-linear over an extended baseline ($0.45<\log{P_0}<2.0$).  Applying a quadratic period-magnitude function capitalizes on the statistical weight of short and long-period Cepheids to improve certain distance determinations. Incidentally, Cepheid distances inferred from 3.6 and 4.5 $\mu$m data are consistent to first-order, thus indicating that CO features present in the 4.5 $\mu$m passband do not seriously compromise the distances evaluated (the 4.5 $\mu$m distances can serve as a consistency check).  The slope and zero-point of the relations appear comparatively insensitive to metallicity variations ($\rm{[Fe/H]}\sim 0$ to -0.75, $|\gamma|<0.1$ mag dex$^{-1}$).  That was determined in part by comparing galaxy distances ($\Delta \rm{[Fe/H]}\sim-0.75$) inferred from the Spitzer period-magnitude relations and NED-D.  The results support prior findings \citep{nk10,fr11}, yet a firm conclusion awaits a comparison between improved data for Cepheids in IC 1613 and the Milky Way.  Further research is likewise required on models, as noted by \citet{ng12b}, with the aim of reproducing the observed period-color trend (Fig.~\ref{fig-lmcpc}).  The latter is non-linear whereby $(3.6-4.5)$ is nearly constant for shorter-period Cepheids, and the color may transition to a bluer convex trough at longer-periods (Fig.~\ref{fig-lmcpm}).  Lastly, an extinction law ($A_{3.6}/E_{B-V}=0.18\pm0.06$, Fig.~\ref{fig-ext}) was derived using Spitzer (GLIMPSE) observations in order to correct the distances for dust obscuration, and to avoid propagating a systematic uncertainty into the cosmic distance scale (and $H_0$).

\subsection*{{\rm \footnotesize ACKNOWLEDGEMENTS}}
\scriptsize{DM is grateful to the following individuals and consortia whose efforts, advice, or encouragement enabled the research: Spitzer, SAGE (M. Meixner, K. Gordon), C. Ngeow, M. Marengo, H. Neilson, CHP (A. Monson, V. Scowcroft, B. Madore, W. Freedman), S. Persson, OGLE (A. Udalski, I. Soszy{\~n}ski), NED-D (I. Steer, B. Madore), GLIMPSE (R. Benjamin), B. Skiff, D. Balam, D. Lane, L. Gallo, R. Thacker, CDS (F. Ochsenbein, T. Boch, P. Fernique), arXiv, and NASA ADS.  WG is grateful for support from the BASAL Centro de Astrofisica y Tecnologias Afines (CATA) PFB-06/2007.}


\begin{thebibliography}{}\setlength{\itemsep}{-1.5mm}
\bibitem[Anderson et al.(2012)]{an12} Anderson, R.~I., Eyer, L., \& Mowlavi, N.\ 2012, arXiv: 1212.5119
\bibitem[Becker(1985)]{be85} Becker, S.\ 1985, IAU Colloq.~82: Cepheids: Theory and Observation, 104
\bibitem[Benedict et al.(2007)]{be07} Benedict, G.~F., McArthur, B.~E., Feast, M.~W., et al.\ 2007, \aj, 133, 1810 
\bibitem[Benjamin et al.(2003)]{be03} Benjamin, R.~A., Churchwell, E., Babler, B.~L., et al.\ 2003, \pasp, 115, 953
\bibitem[Bono et al.(2010)]{bo10} Bono, G., Caputo, F., Marconi, M., \& Musella, I.\ 2010, \apj, 715, 277 
\bibitem[Carraro et al.(2013)]{ca12} Carraro, G., Turner, D., Majaess, D., \& Baume, G.\ 2013, A\&A, in press (arXiv:1305.4309) 
\bibitem[Ferrarese et al.(2007)]{fe07} Ferrarese, L., Mould, J.~R., Stetson, P.~B., et al.\ 2007, \apj, 654, 186
\bibitem[Freedman et al.(2001)]{fr01} Freedman, W.~L., Madore, B.~F., Gibson, B.~K., et al.\ 2001, \apj, 553, 47 
\bibitem[Freedman et al.(2009)]{fr09} Freedman, W.~L., Rigby, J., Madore, B.~F., et al.\ 2009, \apj, 695, 996 
\bibitem[Freedman \& Madore(2011)]{fm11} Freedman, W.~L., \& Madore, B.~F.\ 2011, \apj, 734, 46 
\bibitem[Freedman et al.(2011)]{fr11} Freedman, W.~L., Madore, B.~F., Scowcroft, V., et al.\ 2011, \aj, 142, 192
\bibitem[Freedman et al.(2012)]{fr12} Freedman, W.~L., Madore, B.~F., Scowcroft, V., et al.\ 2012, \apj, 758, 24 
\bibitem[Fouque \& Gieren(1997)]{fg97} Fouque, P., \& Gieren, W.~P.\ 1997, \aap, 320, 799 
\bibitem[Fouqu{\'e} et al.(2007)]{fo07} Fouqu{\'e}, P., Arriagada, P., Storm, J., et al.\ 2007, \aap, 476, 73
\bibitem[Gieren et al.(2005)]{gi05} Gieren, W., Storm, J., Barnes, T.~G., III, et al.\ 2005, \apj, 627, 224
\bibitem[Gieren et al.(2006)]{gi06} Gieren, W., Pietrzy{\'n}ski, G., Nalewajko, K., et al.\ 2006, \apj, 647, 1056 
\bibitem[Gordon et al.(2011)]{go11} Gordon, K.~D., Meixner, M., Meade, M.~R., et al.\ 2011, \aj, 142, 102 
\bibitem[Grieve et al.(1985)]{gr85} Grieve, G.~R., Madore, B.~F., \& Welch, D.~L.\ 1985, \apj, 294, 513
\bibitem[Hackwell \& Gehrz(1974)]{hg74} Hackwell, J.~A., \& Gehrz, R.~D.\ 1974, \apj, 194, 49 
\bibitem[Luck et al.(1998)]{lu98} Luck, R.~E., Moffett, T.~J., Barnes, T.~G., III, \& Gieren, W.~P.\ 1998, \aj, 115, 605 
\bibitem[Macri et al.(2006)]{mac06} Macri, L.~M., Stanek, K.~Z., Bersier, D., Greenhill, L.~J., \& Reid, M.~J.\ 2006, \apj, 652, 1133
\bibitem[Madore(1985)]{ma85} Madore, B.~F.\ 1985, IAU Colloq.~82: Cepheids: Theory and Observation, 166 
\bibitem[Madore et al.(2009)]{ma09} Madore, B.~F., Freedman, W.~L., Rigby, J., et al.\ 2009, \apj, 695, 988 
\bibitem[Majaess et al.(2009)]{maj09} Majaess, D., Turner, D., \& Lane, D.\ 2009, \actaa, 59, 403 
\bibitem[Majaess(2010)]{maj10} Majaess, D.\ 2010, \actaa, 60, 55
%\bibitem[Majaess et al.(2011)]{maj11} Majaess, D.~J., Turner, D.~G., Lane, D.~J., \& Krajci, T.\ 2011, JAAVSO, 39, 219 
\bibitem[Majaess et al.(2012)]{maj12} Majaess, D., Turner, D., Gieren, W., Balam, D., \& Lane, D.\ 2012, \apjl, 748, L9 
\bibitem[Majaess et al.(2013)]{maj13} Majaess, D., Turner, D., Gieren, W., Berdnikov, L., \& Lane, D.\ 2013, \apss, 344, 381 
\bibitem[Marengo et al.(2010)]{mar10} Marengo, M., Evans, N.~R., Barmby, P., et al.\ 2010, \apj, 709, 120 
\bibitem[Meixner et al.(2006)]{me06} Meixner, M., Gordon, K.~D., Indebetouw, R., et al.\ 2006, \aj, 132, 2268
\bibitem[Monson et al.(2012)]{mo12} Monson, A.~J., Freedman, W.~L., Madore, B.~F., et al.\ 2012, \apj, 759, 146 
\bibitem[Mottini et al.(2006)]{mo06} Mottini, M., Romaniello, M., Primas, F., et al.\ 2006, \memsai, 77, 156
\bibitem[Nataf et al.(2013)]{na12} Nataf, D.~M., Gould, A., Fouqu{\'e}, P., et al.\ 2013, \apj, 769, 88 
\bibitem[Neilson et al.(2009)]{ne09} Neilson, H.~R., Ngeow, C.-C., Kanbur, S.~M., \& Lester, J.~B.\ 2009, \apj, 692, 81 
\bibitem[Neilson et al.(2012)]{ne12} Neilson, H.~R., Nardetto, N., Ngeow, C.-C., Fouqu{\'e}, P., \& Storm, J.\ 2012, \aap, 541, A134
\bibitem[Ngeow \& Kanbur(2008)]{nk08} Ngeow, C., \& Kanbur, S.~M.\ 2008, \apj, 679, 76 
\bibitem[Ngeow \& Kanbur(2010)]{nk10} Ngeow, C.-C., \& Kanbur, S.~M.\ 2010, \apj, 720, 626
\bibitem[Ngeow et al.(2012a)]{ng12a} Ngeow, C.-C., Marconi, M., Musella, I., Cignoni, M., \& Kanbur, S.~M.\ 2012 (a), \apj, 745, 104 
\bibitem[Ngeow et al.(2012b)]{ng12b} Ngeow, C.-C., Kanbur, S.~M., Bellinger, E.~P., et al.\ 2012 (b), \apss, 341, 105 
\bibitem[Ngeow et al.(2012c)]{ng12c} Ngeow, C.-C., Neilson, H.~R., Nardetto, N., \& Marengo, M.\ 2012 (c), \aap, 543, A55 
\bibitem[Persson et al.(2004)]{pe04} Persson, S.~E., Madore, B.~F., Krzemi{\'n}ski, W., et al.\ 2004, \aj, 128, 2239 
\bibitem[Pietrzy{\'n}ski et al.(2010)]{pi10} Pietrzy{\'n}ski, G., Thompson, I.~B., Gieren, W., et al.\ 2010, \nat, 468, 542 
\bibitem[Pietrzy{\'n}ski et al.(2011)]{pi11} Pietrzy{\'n}ski, G., Thompson, I.~B., Graczyk, D., et al.\ 2011, \apjl, 742, L20 
\bibitem[Planck Collaboration et al.(2013)]{pl13} Planck Collaboration, Ade, P.~A.~R., Aghanim, N., et al.\ 2013, arXiv:1303.5076 
\bibitem[Romaniello et al.(2008)]{ro08} Romaniello, M., Primas, F., Mottini, M., et al.\ 2008, \aap, 488, 731 
\bibitem[Sandage(1988)]{sa88} Sandage, A.\ 1988, \pasp, 100, 935
\bibitem[Scowcroft et al.(2011)]{sc11} Scowcroft, V., Freedman, W.~L., Madore, B.~F., et al.\ 2011, \apj, 743, 76 
\bibitem[Skiff(2013)]{sk13} Skiff, B.~A.\ 2013, VizieR catalog: General Catalogue of Stellar Spectral Classifications.
\bibitem[Soszy{\~n}ski et al.(2008)]{so08} Soszy{\~n}ski, I., et al.\ 2008, Acta Astronomica, 58, 163 
\bibitem[Soszy{\~n}ski et al.(2010)]{so10} Soszy{\~n}ski, I., Poleski, R., Udalski, A., et al.\ 2010, \actaa, 60, 17
\bibitem[Steer \& Madore(2011)]{sm11} Steer, I., Madore, B. \ 2011, NED-D: NASA/IPAC Extragalactic Database Master List of Galaxy Distances, \url{http://ned.ipac.caltech.edu/Library/Distances/}
\bibitem[Stetson et al.(2004)]{st04} Stetson, P.~B., McClure, R.~D., \& VandenBerg, D.~A.\ 2004, \pasp, 116, 1012 
\bibitem[Strai{\v z}ys \& Lazauskait{\.e}(2009)]{sl09} Strai{\v z}ys, V., \& Lazauskait{\.e}, R.\ 2009, Baltic Astronomy, 18, 19
\bibitem[Storm et al.(2011)]{st11} Storm, J., Gieren, W., Fouqu{\'e}, P., et al.\ 2011, \aap, 534, A94
\bibitem[Tammann et al.(2003)]{ta03} Tammann, G.~A., Sandage, A., \& Reindl, B.\ 2003, \aap, 404, 423 
\bibitem[Tammann \& Reindl(2012)]{tr12} Tammann, G.~A., \& Reindl, B.\ 2012, \apss, 341, 3 
\bibitem[Turner(1989)]{tu89} Turner, D.~G.\ 1989, \aj, 98, 2300 
\bibitem[Turner(2010)]{tu10} Turner, D.~G.\ 2010, Ap\&SS, 326, 219 
\bibitem[Turner(2012a)]{tu12} Turner, D.~G.\ 2012 (a), \apss, 337, 303
\bibitem[Turner(2012b)]{tu12b} Turner, D.~G.\ 2012 (b), JAAVSO, 40, 502 
\end{thebibliography}
\end{document}